\documentstyle[aaspp4,11pt]{article}

\newcommand{\be}{\begin{equation}}
\newcommand{\ee}{\end{equation}}
\newcommand{\bea}{\begin{eqnarray}}
\newcommand{\eea}{\end{eqnarray}}
\newcommand{\etal}{{\it et al.}}

\begin{document}

\title{Interplanetary Measures Can Not Bound the Cosmological Constant}
\author{Edward L. Wright}
\affil{UCLA Astronomy \& Astrophysics Division}
\authoraddr{PO Box 951562, Los Angeles, CA 90095-1562}

\begin{abstract}
The effect of a cosmological constant on the precession of the line of 
apsides is ${\cal O}(\Lambda c^2 r^3/GM)$ which is
$3(H_\circ P)^2/8\pi^2 \approx 10^{-23}$ for a vacuum-dominated 
Universe with Hubble constant
$H_\circ = 65$~km/sec/Mpc and for the orbital period $P = 88$~days
of Mercury.  This is unmeasurably small, so
planetary perturbations cannot be used to limit the cosmological
constant, contrary to the suggestion by Cardona \& Tejeiro (1998).
\end{abstract}


\section{Introduction}

Cardona \& Tejeiro \markcite{CT98} (1998) maintain that the 
precession of the perihelion
of Mercury can be used to set limits on the cosmological constant that are
within a factor of 10-100 of the limits set using cosmological observations.
Since the effect of the cosmological constant is only expected to be
significant at large radii, this result is quite surprising.  In this note I
show that it is incorrect.

\section{Calculation}

I use the same Gibbons \& Hawking \markcite{GH77} (1977) metric 
used by Cardona \& Tejeiro \markcite{CT98} (1998):
\be
ds^2 = 
B(r) c^2 dt^2 - A(r) dr^2 -r^2\left(d\theta^2 + \sin^2\theta d\phi^2\right)
\ee
where
\be
B(r) = 1 - \frac{2GM}{c^2r} - \frac{\Lambda r^2}{3}, \quad\quad A = B^{-1}.
\ee
The angular distance between the perihelion $(r_-)$ and aphelion $(r_+)$
is given by
\be
\phi(r_+) - \phi(r_-) = \int_{r_-}^{r_+} \frac{A(r)^{1/2} dr}
{r^2\left[J^{-2} \left(B^{-1}(r) - E\right) -r^{-2}\right]^{1/2}}
\ee
where $J$ and $E$ are integrals of the motion,
and the precession of the perihelion per orbit is
\be
\Delta\phi = 2\left[\phi(r_+) - \phi(r_-)\right] - 2\pi.
\ee
Unlike Cardona \& Tejeiro \markcite{CT98} (1998) I use Eqn(8.6.3) from 
Weinberg \markcite{W72} (1972) to compute
$\Delta\phi$:
\be
\phi(r_+) - \phi(r_-) = \int_{r_-}^{r_+} \frac{A(r)^{1/2}}{r^2}
\left[\frac{r_-^2(B^{-1}(r)-B^{-1}(r_-))-r_+^2(B^{-1}(r)-B^{-1}(r_+))}
{r_+^2 r_-^2(B^{-1}(r_+)-B^{-1}(r_-))}-\frac{1}{r^2}\right]^{-1/2} dr
\label{eq:dphi}
\ee
The function inside the $[\;]$ in Eqn (\ref{eq:dphi}) vanishes at $r_+$ and
$r_-$, so I can write
\be
\left[\frac{r_-^2(B^{-1}(r)-B^{-1}(r_-))-r_+^2(B^{-1}(r)-B^{-1}(r_+))}
{r_+^2 r_-^2(B^{-1}(r_+)-B^{-1}(r_-))}-\frac{1}{r^2}\right]
\approx C\left(\frac{1}{r}-\frac{1}{r_-}\right)
\left(\frac{1}{r}-\frac{1}{r_+}\right)
\label{eq:C}
\ee
and this will be a good approximation for slightly eccentric orbits
with any $\Lambda$ and for any eccentricity at zero $\Lambda$.
Unlike the PPN case worked out by Weinberg \markcite{W72} (1972)
I can not evaluate $C$
by going to $r = \infty$ because the $\Lambda$ term diverges there.
I find $C$ by taking the second derivative of Eqn (\ref{eq:C}) with
respect to the variable $u = 1/r$.
Noting that $B^{-1} = A$ for this metric, the constant $C$ is given by
\be
C = \frac{(u_+-u_-)(u_-+u_+)A^{\prime\prime}(u)}
{2 (u_+-u_-)A^\prime(u)} - 1  = -1 +
\left.\frac{u A^{\prime\prime}(u)}{A^\prime(u)}
\right\vert_{u = L^{-1}}
\ee
where $L^{-1} = 0.5(u_+ + u_-)$ and $L = a(1-e^2)$ is the
{\em semilatus rectum} of the elliptical orbit.  In this formula
I have approximated $A(u_+)-A(u_-)$ by 
$(u_+-u_-)A^\prime((u_++u_-)/2)$, which will also be a
good approximation for slightly eccentric orbits
with any $\Lambda$ and for any eccentricity at zero $\Lambda$.
% will be a good approximation
% for small eccentricities and ${\cal O}(M^3)$ for zero $\Lambda$.
Expanding $B^{-1}$ gives
\be
A = B^{-1} 
= 1 + r_s u + r_s^2 u^2 + \frac{\Lambda}{3u^2} + \frac{2\Lambda r_s}{3u} +
\ldots
\ee
where the Schwarzschild radius is $r_s = 2GM/c^2$,
so
\be
A^\prime = r_s + 2 r_s^2 u - \frac{2\Lambda}{3u^3} - \frac{2\Lambda r_s}{3u^2}
\ee
and
\be
A^{\prime\prime} = 2 r_s^2 + \frac{2\Lambda}{u^4} +\frac{4\Lambda r_s}{3u^3}
\ee
Therefore, 
\bea
C & = & \frac{2 r_s^2 u + 2\Lambda u^{-3} + (4/3)\Lambda r_s u^{-2}}
{r_s + 2 r_s^2 u - (2/3)\Lambda u^{-3} - (2/3)\Lambda r_s u^{-2}} 
- 1\nonumber\\
& = & 
\frac{- r_s + (8/3)\Lambda u^{-3} +2 \Lambda r_s u^{-2}}
{r_s + 2 r_s^2 u - (2/3)\Lambda u^{-3}-(2/3)\Lambda r_s u^{-2}} \nonumber \\
& \approx & - \left(1 - 2 r_s u - \frac{2\Lambda}{r_s u^3} 
-\frac{4\Lambda}{3u^2}\right)
\eea
Then the integral for $\phi$ becomes
\be
\phi(r_+) - \phi(r_-) = \int_{u_-}^{u_+} \frac{A(u)^{1/2} du}
{\left[C(u-u_+)(u-u_-)\right]^{1/2}}
= \pi \left(1 + \frac{3}{2} r_s u + \frac{5}{6}\Lambda u^{-2} 
+ \frac{\Lambda}{r_s u^3}\right)
\ee
and the precession per orbit is
\be
3\pi r_s u + \Lambda u^{-2}\left(\frac{2\pi}{r_s u} + \frac{5\pi}{3}\right)
\ee
evaluated at $u = L^{-1}$.
The first term is the standard GR perihelion
precession of $6\pi GM/(L c^2)$.
Since $r_s u << 1$ in the Solar System, the additional precession 
due to the cosmological constant is
\be
\Delta\phi_\Lambda = \frac{2\pi\Lambda c^2 L^3}{GM}
= 6\pi \frac{\rho_{vac}}{\overline{\rho}}\;\mbox{rad/orbit}
\ee
where $\overline{\rho}$ is the average density within a sphere of
radius $L$, and $\rho_{vac} = \Lambda c^2/(8\pi G)$ is the 
vacuum density equivalent of the cosmological constant.

\section{Discussion}

For Mercury with $r_s u = 5 \times 10^{-8}$
\be
\Delta\phi_\Lambda = \frac{2\pi\Lambda L^3}{r_s} = 1.3 \times 10^8 \Lambda L^2
\;\mbox{rad/orbit}
\ee
Since the uncertainty in the precession of the perihelion of Mercury is about
$0.1^{\prime\prime}$ per century or $10^{-9}$ rad/orbit, the limit obtained
on the cosmological constant is
\be
\Lambda < 10^{-32.5}\;\mbox{km$^{-2}$}
\ee
which is $10^{12.5}$ times weaker than the result of 
Cardona \& Tejeiro \markcite{CT98} (1998).
The corrected limit is
not competitive with the cosmological limit of
$\Lambda < 10^{-46}\;\mbox{km$^{-2}$}$ (Kochanek \markcite{K96} 1996).
More distant planets would give better limits: 
$\Lambda = 10^{-32.5}\;\mbox{km$^{-2}$}$ would induce a 
$100^{\prime\prime}$ per
century precession in the perihelion of Pluto, since the precession
per century scales like $L^3/a^{3/2} \approx a^{3/2}$.
Hellings \markcite{H84} (1984) determined the PPN parameter $\beta$ to
an accuracy $\Delta\beta = 10^{-3}$ from Mars data including Viking
lander ranging, and this translates into an uncertainty on the cosmological
constant of
\be
\Delta\Lambda = -\frac{1}{2} r_s^2 L^{-4} \Delta\beta < 
10^{-35.7}\;\mbox{km$^{-2}$}
\ee

The precession rate of the perihelion is 3 times larger than the
change in the mean motion for circular orbits
\be
\frac{\Delta n}{n} = -\frac{\Lambda r^3}{3 r_s}
= -\frac{\rho_{vac}}{\overline{\rho}}
\ee
used by
Anderson \etal \markcite{ALKDD95}\ (1995) to search for dark matter
in the Solar System and does not require
an independent determination of the distance to the planet,
so the precessions of the perihelia of the planets provide
the most sensitive Solar System test for a cosmological constant.  
The limit on $\Lambda$ from Mars observations is two orders
of magnitude better than the limit based on
the Anderson \etal \markcite{ALKDR95}\ (1995) result from the
mean motion of Neptune.  But no Solar System test can ever compete 
with tests based on the large scale geometry of the Universe.

\acknowledgements
This work was supported in part by NASA grant NAG 5-1309 to UCLA.

\end{document}